\documentclass[12pt,twoside]{article}
\usepackage{xrb2007}
\usepackage{psfig}

\begin{document}

% select your session by uncommenting the appropriate line
%\session{Jets}
%\session{Jet and Black Hole Binaries}
\session{Faint Galactic XRB Populations}

%\session{Faint XRBs and Galactic LMXBs}
%\session{Obscured XRBs and INTEGRAL Sources}
%\session{ULXs}
%\session{Extragalactic Populations}
%\session{Future Missions and Surveys}
%\session{Population Synthesis}

\shortauthor{Heinke et al.}
\shorttitle{CVs in Globulars and the Galaxy}

\title{Cataclysmic Variables in Globular Clusters, the Galactic Center, and Local Space}
\author{Craig O. Heinke}
\affil{University of Virginia, Astronomy Dept., PO Box 400325, Charlottesville VA 22903; cheinke@virginia.edu}

\author{Ashley J. Ruiter}
\affil{Dept. of Astronomy, New Mexico State University, 1320 Frenger Mall, Las Cruces, NM 88003}

\author{Michael P. Muno}
\affil{Space Radiation Laboratory, California Institute of Technology, Pasadena, CA 91125}

\author{Krzysztof Belczynski}
\affil{Dept. of Astronomy, New Mexico State University, 1320 Frenger Mall, Las Cruces, NM 88003; Tombaugh Fellow}

\begin{abstract}
We compare the X-ray spectra and luminosities, in the 2-8 keV band, of known and suspected cataclysmic variables (CVs) in different environments, assessing the nature of these source populations.  These objects include nearby CVs observed with ASCA; the Galactic Center X-ray source population identified by Muno et al.; and likely CVs identified in globular clusters.  Both of the latter have been suggested to be dominated by magnetic CVs.  We find that the brighter objects in both categories are likely to be magnetic CVs, but that the fainter objects are likely to include a substantial contribution from normal CVs.  The strangely hard spectra observed from the Galactic Center sources reflect the high and variable extinction, which is significantly greater than the canonical $6\times10^{22}$ cm$^{-2}$ over much of the region, and the magnetic nature of many of the brightest CVs.  The total numbers of faint Galactic Center sources are compatible with expectations of the numbers of CVs in this field. 
\end{abstract}

\section{Introduction}

The unprecedented spatial resolution of the {\it Chandra X-ray 
Observatory} allows us to study populations of faint X-ray sources 
at distances of kiloparsecs.
Large numbers of X-ray sources of moderate luminosities 
($10^{31}<L_X<10^{34}$ ergs/s) have been discovered in several 
Galactic environments (e.g. star-forming regions, globular clusters, 
the Galactic Center) in recent years.  
For instance, \citet{Grindlay01a,Pooley02a} and others have found large numbers 
of X-ray sources in globular clusters, many of which are optically 
identified as CVs.  
\citet{Muno03} identified a large population of $\sim$2000 faint, 
spectrally hard X-ray sources
associated with the central 40 pc around Sgr A*. 
 
Interpretation of these observations in terms of known categories of 
X-ray sources is sometimes hampered by the different observing bands 
and models used for studies of nearby, well-known sources versus these 
distant populations.  
The Galactic Center sources have been inferred \citep{Muno03,Muno04a},
 by their X-ray luminosity and spectral hardness, to be composed 
largely of the class of CVs known as intermediate polars \citep{Patterson94}. 
Intermediate polars have white dwarf accretors with relatively high 
magnetic fields, which force the accreting material to follow the 
magnetic field lines of the accretor down onto the magnetic poles. 
The brightest CVs in globular clusters have been argued to be too 
X-ray luminous for normal CVs \citep{Verbunt97}, and several pieces of 
evidence point to a magnetic nature of at least some CVs in globular 
clusters, including He II $\lambda$4686 emission observed in NGC 6397 
CVs \citep{Grindlay95}, X-ray periodicities in some of the brightest 
CVs in 47 Tuc \citep{Grindlay01a},
enhanced $N_H$ columns towards the brightest CVs \citep{Heinke05a}, 
 and the lack of dwarf nova outbursts 
from globular clusters \citep{Shara96,Edmonds03b,Dobrotka06}.  

However, the question of whether magnetic CVs are more common than normal 
CVs in globular clusters or the Galactic Center has not yet been explored.  
We have directly compared, for the first time, the spectra of Galactic Center X-ray sources, 
and likely globular cluster CVs (those with optical counterparts), with archival ASCA X-ray spectra of well-known nearby CVs.  Full details will be published in Heinke et al. (2008, in prep).

\section{Local CVs}

We cross-correlated the CV database of \citet{Ritter03}
\footnote{http://physics.open.ac.uk/RKcat/} with the ASCA
observations in the HEASARC archive.  We chose 20 confirmed IPs with 
substantial pointed ASCA observations, 11 polars, 8 novalike (or outbursting dwarf nova) CVs, and 11 quiescent dwarf nova observations \citep[see ][]{Baskill05,Ezuka99}.  We used the archived spectra and responses from the ASCA archive, 
and extracted appropriate background spectra. 

 We fit the ASCA spectra from 2-8 keV, in order to compare the same energy range for all our data.  A power-law model with a single gaussian (representing the 
combination of Fe lines) generally gave acceptable fits to the data.  We find (omitting 3 CVs with poorly determined indices) a clear difference between the fitted photon indices of the magnetic (mean $\Gamma=1.22$, $\sigma=0.33$) and the nonmagnetic (mean $\Gamma=1.97$, $\sigma=0.20$) ASCA CVs.  We find consistency, however, between the average IP and polar spectra, and between the average quiescent dwarf nova spectrum and the average novalike/outbursting dwarf nova spectrum.  High-$B$ IPs and polars (e.g. AE Aqr and V884 Her) have suppressed hard X-ray radiation, producing low $L_X$s and photon indices.  Excluding these high-$B$ systems, we do not find strong dependence of the photon index with $L_X$ within either the magnetic CVs or nonmagnetic CVs as groups.  

\section{Globular Cluster CVs}

For comparison with the ASCA observations and Galactic Center sources, we perform homogeneous X-ray spectral fitting of 23 (optically) identified CVs in globular clusters (47 Tuc, NGC 6397, $\omega$ Cen, NGC 6752, and M4), with $>$90 counts above 2 keV.  We compare the fitted power-law photon indices and estimated X-ray luminosities of the globular cluster CVs with the observed CV populations in Figure 1a.  The locations of globular cluster X-ray sources in this plot suggest that some (the brightest and hardest) are likely magnetic systems, while many others are likely nonmagnetic systems.  

\begin{figure}
\psfig{figure=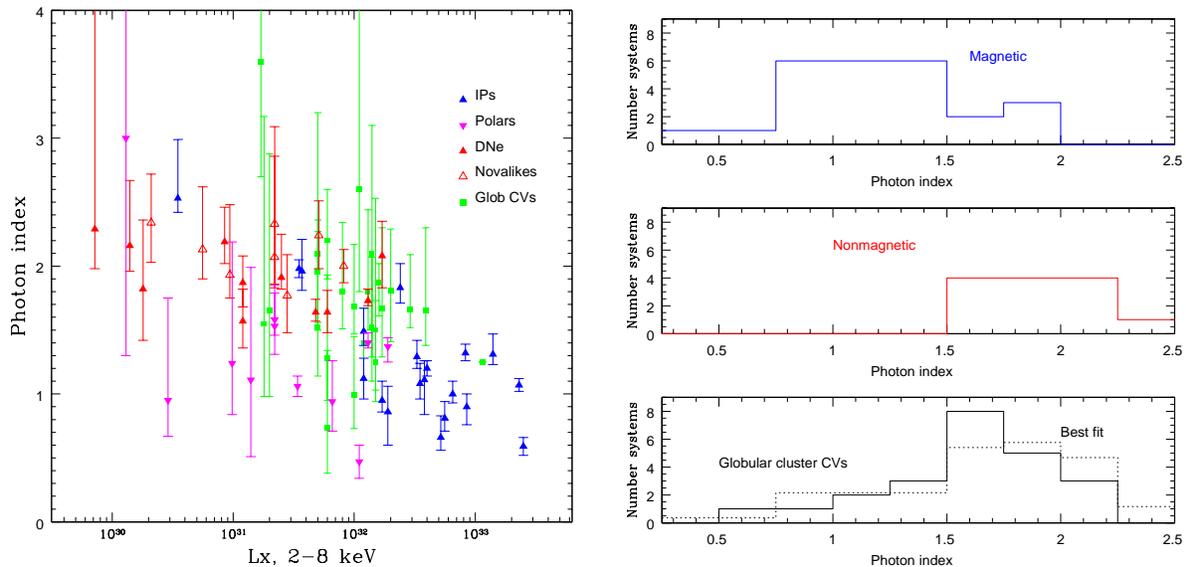,width=6.4in}
\caption{{\it Left:} X-ray luminosities vs. photon index measurements for 
various classes of nearby CVs vs. those measured for globular cluster X-ray 
sources optically identified as CVs.  
{\it Right:} Histograms of the photon indices of magnetic, nonmagnetic, and globular cluster CVs.  The bottom panel also shows the best-fit distribution of photon index produced by scaling the histograms of magnetic and nonmagnetic CV photon indices to match the globular cluster CVs.
}
\end{figure}

We selected an X-ray luminosity range ($10^{31}$ to $2\times10^{33}$ ergs/s) that includes all globular cluster CVs in our list.  In this range we find 25 magnetic systems and 12 nonmagnetic systems with ASCA spectra.  We show histograms of the photon indices of magnetic, nonmagnetic, and cluster CVs in Fig. 1b.  
We scaled the histograms of magnetic and nonmagnetic CV indices to match the histogram of globular cluster CV indices. We find a best fit of 39$^{+12}_{-15}$($1\sigma$)\% magnetic systems, i.e. 5 to 12 of the 23 confirmed globular cluster CVs, with the rest being nonmagnetic systems.  Since these systems have been identified in a nonuniform way, with strong X-ray selection, it is likely that the fraction of magnetic CVs in globular clusters is lower than this value.  We do not find evidence that the fraction of magnetic CVs (polars and intermediate polars) in globular clusters is higher than the $\sim10$\% estimated for the field \citep{Liebert03}.

\section{Galactic Center Sources}

\citet{Muno04a} characterize the Galactic Center sources, finding typical X-ray 
luminosities of $3\times10^{31}$--$10^{33}$ ergs/s, and very hard X-ray spectra, with equivalent photon indices generally between 1 and -1.  This is rather harder than the typical globular cluster CV or field CV, or even the magnetic CVs in either location \citep{Heinke06b}.  \citet{Muno04a} pointed out that selection effects (the high extinction and diffuse background) probably had a role in the hardness of these sources.  To test whether the Galactic Center sources were consistent with a combination of magnetic and nonmagnetic CVs, or even with just magnetic CVs, we undertook MARX simulations in which we added sources of known properties to 414 ks (2/3 of the total used by \citet{Muno03}; the 3 longest observations, to reduce computing time) of the real {\it Chandra} observations of the Galactic Center, ran detection algorithms and measured the colors of the detected fake sources.  

For our simulated source population, we choose a population synthesis model using the StarTrack code \citep{Belc08} as implemented in \citet{Ruiter06}, with updates to compute the X-ray luminosities of magnetic and nonmagnetic CVs using the prescription of \citet{Patterson85}.  We show the luminosity functions for magnetic and nonmagnetic systems in Figure 2.  We assume that 10\% of the total CV population are magnetic systems.  

\begin{figure}
\psfig{figure=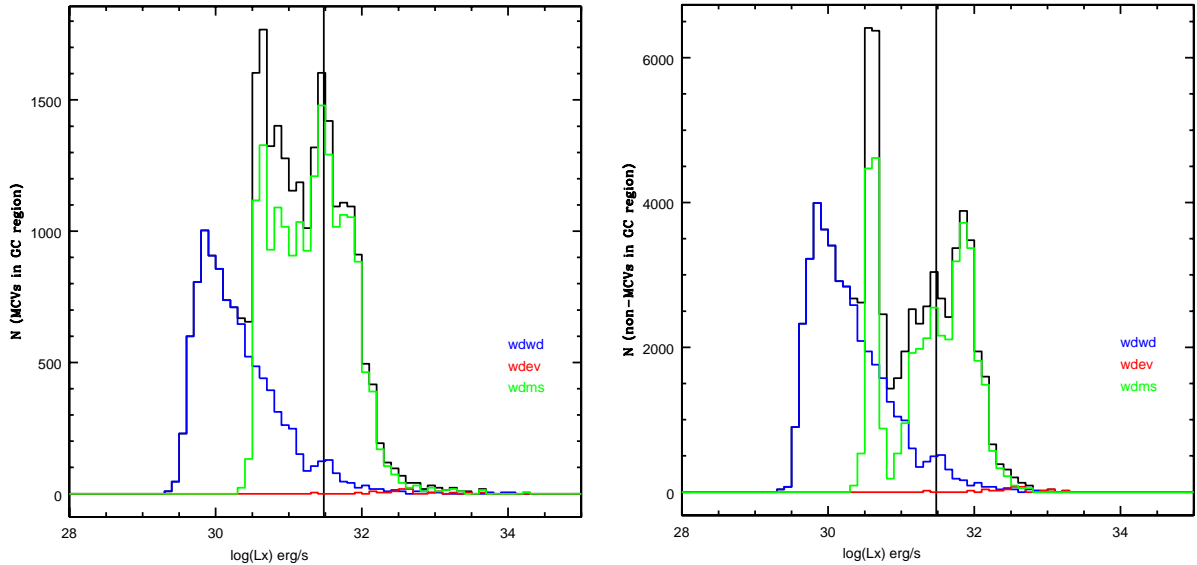,width=6.4in}
\caption{{\it Left:}  Histograms of $L_X$ (2-8 keV) from StarTrack population synthesis
for magnetic CVs.  Blue: white dwarf--white dwarf systems, green: white dwarf--main sequence systems, red: white dwarf--evolved star systems.  Black vertical line: rough lower limit of Muno observations. 
{\it Right:} Same as left, but for nonmagnetic CVs.  
}
\end{figure}

For the spectra of the simulated CVs, we choose absorbed power-law spectra with single gaussians to represent the Fe K line complex, with average energies and equivalent widths based on the ASCA fits.  Our absorption includes both photoelectric absorption (using the XSPEC model {\it phabs}) and scattering by dust (using P. Predehl's XSPEC model {\it scatter}, \citet{Predehl03}).  For nonmagnetic systems, we use an average photon index of 1.97.  For magnetic systems, we produce simulated systems with photon indices of 0.72, 1.22, and 1.72, with a distribution set by the results from the ASCA magnetic CV spectra.  The only parameter we adjust to match the observations is the extinction.  We compare the simulated systems to the real sources, extracted in the same way from the same Galactic Center data.  We compare the medium and hard colors \citep[defined by ][as (h-s)/(h+s), where the medium color uses the 2.0-3.3 and 3.3-4.7 keV bands, and 3.3-4.7 and 4.7-8 for the hard color]{Muno03} and measured photon fluxes (see Figure 3).

\begin{figure}
\psfig{figure=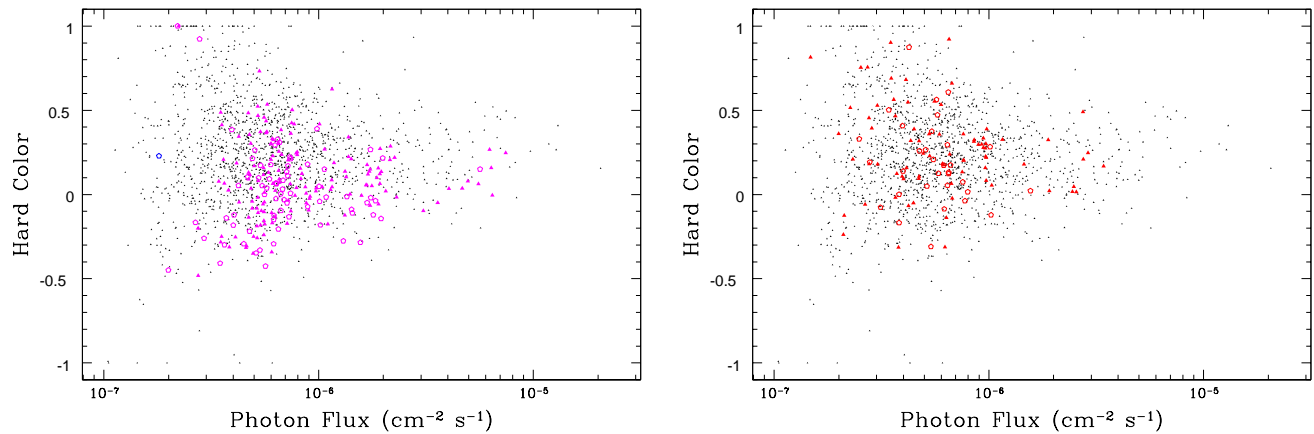,width=7in}
\caption{  Photon flux vs. hard color for Galactic Center sources 
(black), and for our simulations with $N_H=6\times10^{22}$ (left, magenta) or $10^{23}$ cm$^{-2}$ (right, red). 
Definitions of photon flux and hard color are the same as in \citet{Muno04a}. 
Triangles indicate simulations of magnetic systems, open pentagons nonmagnetic systems. 
}
\end{figure}

\begin{figure}
\psfig{figure=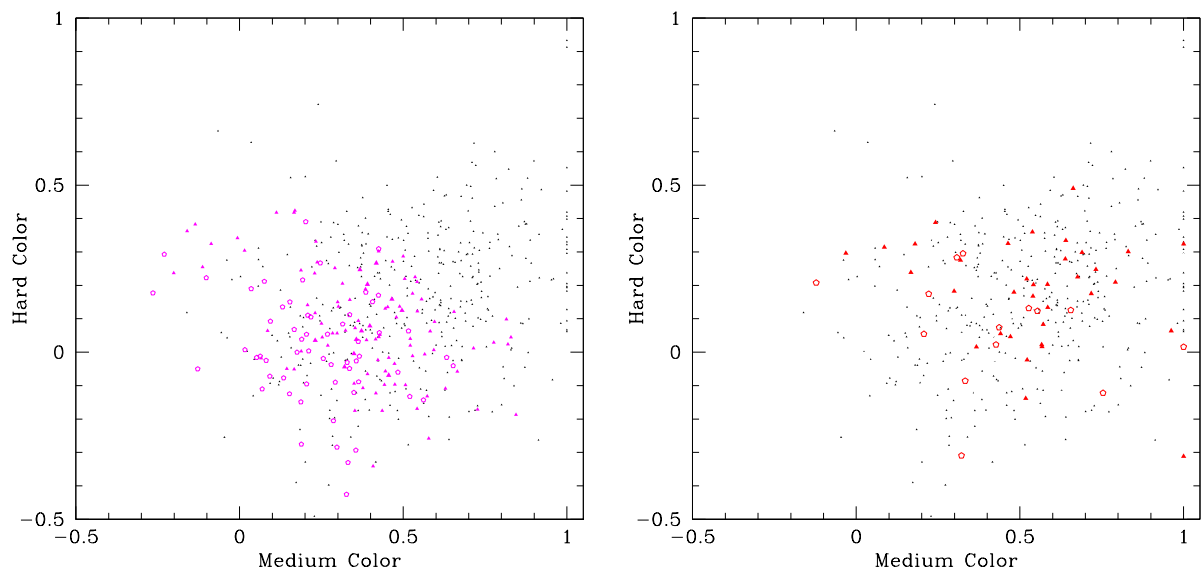,width=6.4in}
\caption{ Medium color vs. hard color for Galactic Center sources (black) 
and for our simulations with $N_H=6\times10^{22}$ (left, magenta) or $10^{23}$ cm$^{-2}$ (right, red).   
Definitions of colors as in \citet{Muno03}; triangles indicate simulations of magnetic systems and open pentagons nonmagnetic systems.
}
\end{figure}

 We find that the real data can be reasonably described with our model only if a higher $N_H$ of $10^{23}$ cm$^{-2}$ (plus dust scattering) absorbs the majority of the sources, rather than the canonical $N_H=6\times10^{22}$.  
Figures 3 (color-flux) and 4 (color-color) show samples of our results, using only $N_H=6\times10^{22}$ (left panels) or  $10^{23}$ cm$^{-2}$ (right panels).  It can be seen that the right panels exhibit much better qualitative matches to the data.  
Similar agreement can be reached using broader ranges of $N_H$ with an average extinction of $10^{23}$ cm$^{-2}$.  

%With this change, the color-color and color-flux diagrams are qualitatively very well modeled, considering the large uncertainties in many aspects of the simulations.  

\begin{figure}
\psfig{figure=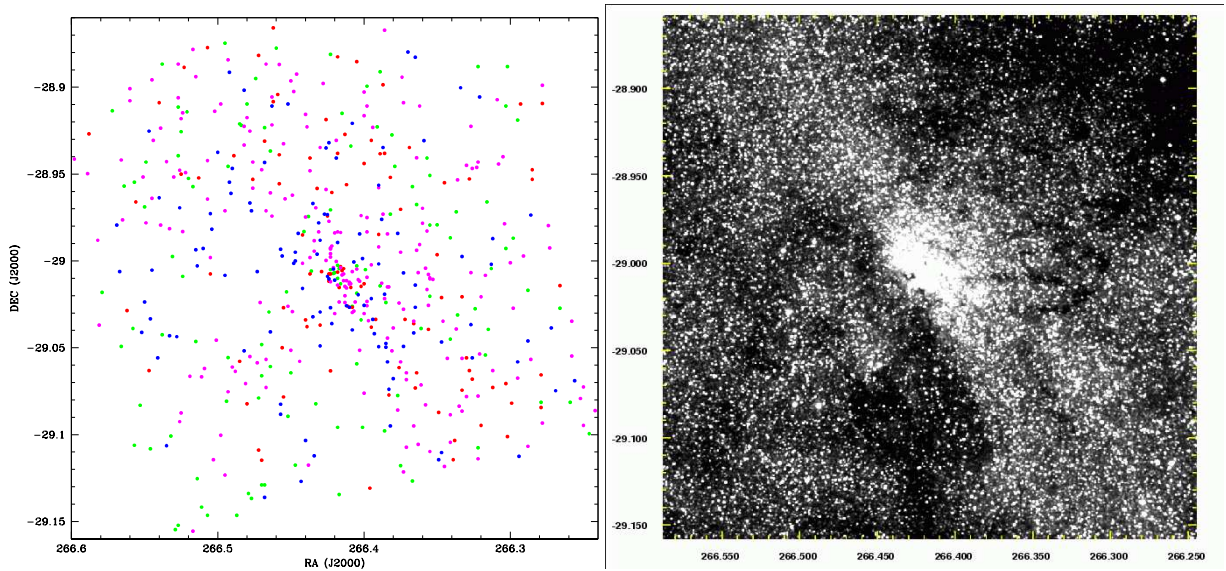,width=6.4in}
\caption{{\it Left:} Locations of Chandra X-ray sources from \citet{Muno04a}, 
coded by the $N_H$ value in fits to thermal plasma spectra.  Green: $N_H<6\times10^{22}$,
red: $6-10\times10^{22}$, magenta: $10-20\times10^{22}$, blue: $>20\times10^{22}$. 
{\it Right:} K-band image of Galactic Center, from the 2MASS survey.  
}
\end{figure}

The interstellar absorption towards the Galactic Center is known to be inhomogeneous and filamentary.  The variations in this absorption have a strong effect on the numbers and hardness of the galactic center X-ray sources seen at different positions.  This can be qualitatively seen in Figure 5, which compares a K-band image of the Galactic Center from the 2MASS survey with the positions and fitted $N_H$ values of Galactic Center X-ray sources.  
Current near-IR observations of the Galactic Center field \citep[e.g. ][]{Gosling06} will improve our understanding of the effects of extinction upon X-rays from the Galactic Center.

The total number of CVs in the Galactic Center may be inferred from our results (with caveats, particularly the variability of extinction in the region).  The total number of simulated CVs required to match the numbers of observed sources, using a single extinction of $N_H=10^{23}$ cm$^{-2}$, is about 7000.  Following \citet{Muno03}, and using an estimate of $1\times10^{-5}$ CVs pc$^{-3}$ \citep{Grindlay05} in local space, we estimate a total CV number in the Galactic Center of 5000.  This numerical agreement indicates that CVs are indeed the major contributor to the Galactic Center X-ray sources, and suggests that these CVs are not significiantly different in their X-ray properties or formation mechanisms from CVs in our galactic neighborhood.

\section{Conclusions}

Studies of faint ($10^{31}<L_X<10^{34}$ ergs/s) hard X-ray sources in globular clusters and the Galactic Center have identified them as primarily CVs, on both observational and theoretical (lack of a sufficiently numerous alternative population) grounds.  They have also suggested that many or most of them are magnetic systems, particularly intermediate polars.  We have compared the spectra of nearby CVs, observed with ASCA, to the low-count spectra or colors of globular cluster CVs and Galactic Center X-ray sources.  We find that significant (although poorly constrained) fractions of the observed cluster and galactic center populations are consistent in their X-ray spectra and fluxes with nonmagnetic CVs.  For the Galactic Center, we require a somewhat higher average extinction than typically assumed ($N_H=10^{23}$ instead of $6\times10^{22}$ cm$^{-2}$). 
We do not find evidence for significant differences between the X-ray properties and fraction of magnetic systems of CVs in local space, vs. those of CVs in globular clusters or the Galactic Center.

\begin{quote}
\verb''\acknowledgements We thank R.~E. Taam and Koji Mukai for useful conversations.   COH has been supported by the Lindheimer Fellowship at Northwestern University, and Chandra grant G07-8078X at the Univ. of Virginia while doing this work.  MPM has been supported by a Hubble Fellowship, while KB has been supported by a Tombaugh Fellowship.  AJR was supported by a Chandra Theory Grant and is thankful to the Dept. of Physics and Astronomy at Northwestern University for their hospitality.  This publication makes use of data products from the Two Micron All Sky Survey, which is a joint project of the University of Massachusetts and the Infrared Processing and Analysis Center, funded by the National Aeronautics and Space Administration and         the National Science Foundation.
\end{quote}

%\bibliographystyle{apj}
%\bibliography{src_ref_list}

\end{document}